\documentclass[
  aps,
  prl,
  reprint,
  superscriptaddress
]{revtex4-2}

\usepackage{amsmath,amssymb,bm}
\usepackage{graphicx}
\usepackage{xcolor}

\begin{document}

\title{An internal mode shifts the depinning threshold of dissociated
dislocations without changing universality}

\author{Mikko J. Alava}
\email[Corresponding author:~]{mikko.alava@aalto.fi}
\affiliation{Department of Applied Physics, Aalto University,
P.O. Box 15600, 00076 Aalto, Espoo, Finland}

\author{Tero J. M\"akinen}
\affiliation{Department of Applied Physics, Aalto University,
P.O. Box 15600, 00076 Aalto, Espoo, Finland}

\begin{abstract}
Driven elastic manifolds exhibit a depinning transition whose large-scale
behavior is governed by a nonequilibrium critical fixed point. Real
interfaces, however, often possess internal degrees of freedom, raising the
question of whether such modes alter depinning universality or only
nonuniversal properties of the transition. We address this question for a
dissociated dislocation, whose two Shockley partials define a fluctuating
stacking-fault width driven by correlated random-alloy stresses. Comparing
exact thresholds on identical disorder realizations with the width either
free or frozen, we find that width fluctuations leave the depinning
universality class unchanged: the roughness exponent $\zeta$ and
correlation-length exponent $\nu$ remain the same even when the width becomes
strongly fluctuating and locally multistable. The threshold, however, shifts.
The shift is governed by the elastic crossover length of the internal mode
rather than by multistability, and can have either sign. An infinite-range mean-field
treatment misses this mechanism. With the physical two-partial elastic kernel
and disorder calibrated against measured strengths, width fluctuations raise
the athermal critical resolved shear stress by 10--35\% over that of a rigid
dissociated dislocation. An internal mode can thus be irrelevant at the
depinning fixed point while remaining quantitatively important for where the
transition occurs.
\end{abstract}

\maketitle

The depinning transition of an elastic manifold in quenched disorder is a
canonical nonequilibrium critical phenomenon~\cite{Nattermann1992,NarayanFisher1993,Fisher1998,Chauve2001,LeDoussal2002,Ferrero2021,wiese2022theory}.
At a sample-dependent threshold $\sigma_c$ a pinned state loses stability,
while near threshold the diverging correlation length, interface geometry,
and dynamics are governed by universal scaling
laws~\cite{Chauve2001,LeDoussal2002,Leschhorn1997}. For long-range elasticity
the same description applies to contact
lines~\cite{ErtasKardar1994Contact,Rosso2002}, crack
fronts~\cite{Schmittbuhl1995,Ramanathan1997} and
dislocations~\cite{Zapperi2001,Bako2008}.  This separation between a
nonuniversal threshold and universal critical behavior is central to
depinning: microscopic disorder determines where the transition occurs,
whereas the long-wavelength fixed point determines how it is approached.
The standard theory, however, describes the driven object by a single
displacement field.  Many physical interfaces possess additional internal
degrees of freedom, and whether these modes renormalize only microscopic
quantities or become relevant variables of the depinning transition remains
largely unresolved.

Internal structure can affect driven interfaces in qualitatively different
ways.  Transverse fluctuations of a driven flux line leave its longitudinal
critical behavior unchanged in an isotropic medium, whereas anisotropy can
generate different scaling~\cite{ErtasKardar1994Flux}.  Magnetic domain walls
provide a contrasting example: internal magnetization supports Bloch
lines~\cite{Herranen2015,Herranen2017}, whose proliferation changes the
driven dynamics~\cite{Herranen2017,HonkanenLaurson2026} and has been
associated with disorder-dependent effective depinning
exponents~\cite{Skaugen2022}.  These examples raise a general question: when
a driven object contains a fluctuating internal coordinate, does that
coordinate change the depinning fixed point, merely shift the critical
stress, or do both?  The two effects are logically distinct but are rarely
separated in the same system.

Dissociated dislocations provide a particularly clean realization of this
problem.  In a face-centered-cubic (fcc) crystal a perfect dislocation
separates into two Shockley partials bounding a stacking fault.  Their mean
displacement defines the translation of the dislocation, while their local
separation $W(x)$ is a spatially resolved internal degree of freedom.  In a
random alloy the two partials experience different quenched stresses, whose
variance~\cite{GeslinRodney2021} and anisotropic
correlations~\cite{GeslinRidaRodney2021} follow from continuum elasticity and
whose consequences for pinning have been studied for single
lines~\cite{Rida2022,ZaiserWu2022}, whereas their elastic interaction and the
stacking-fault energy restore their separation.  The width can therefore
fluctuate strongly and even possess multiple metastable local
configurations~\cite{Vaid2022}.  Molecular-dynamics simulations of
multicomponent solid solutions indeed show strong dislocation roughening and
a broad range of apparent depinning exponents~\cite{Esfandiarpour2022}.  At
the same time, the issue is directly relevant to strengthening: established
solute-strengthening theories treat the dissociated dislocation as a rigid
object~\cite{Leyson2010,Varvenne2016}.

Here we disentangle the universal and nonuniversal consequences of this
internal mode.  We study two interacting elastic lines driven through
spatially correlated disorder derived from random-alloy stress statistics,
locate the threshold of every realization exactly, and compare it with a
control in which the stacking-fault width is frozen.  The width is a massive
mode with an elastic crossover length $\ell^\ast$, but disorder can
nevertheless make it strongly fluctuating and locally multistable.  We find
that freeing this mode does not change the depinning universality class: the
roughness exponent $\zeta$ and correlation-length exponent $\nu$ agree with
those of the frozen-width problem throughout the explored regime.  In
contrast, the threshold shifts, in either direction.  The shift is controlled
by $\ell^\ast$ itself, not by the onset of local multistability or by the
composition of the instability that triggers depinning. Consequently an
infinite-range treatment gives both the wrong magnitude and, in part of the
parameter range, the wrong sign.  Finally, using the physical elastic kernels
of two Shockley partials and disorder strengths appropriate to concentrated
fcc alloys, we find that width fluctuations increase the athermal critical
resolved shear stress (CRSS) by 10--35\%.\\

\emph{Model}---%
We represent the Shockley partials by single-valued displacements
$y_1(x,t)$ and $y_2(x,t)$ along the glide direction (where $x$~is the coordinate perpendicular to the glide direction and $t$ time).  In units in which
their bare mobilities are one, their overdamped dynamics are
\begin{align}
\partial_t y_1 &=
\mathcal{J}[y_1]+\sigma
+\gamma\left(\frac{W_0}{W}-1\right)
+F_1(x,y_1), \label{eq:1}\\
\partial_t y_2 &=
\mathcal{J}[y_2]+\sigma
-\gamma\left(\frac{W_0}{W}-1\right)
+F_2(x,y_2), \label{eq:2}
\end{align}
where $\mathcal{J}$ is the self-interaction of a single partial, $W(x,t)=y_1-y_2$, $W_0$ is its disorder-free equilibrium value,
and $\gamma$ is the stacking-fault stress.  
Each term is a glide force per unit length divided by the partial Burgers vector $b_p$, i.e.~corresponds to a resolved shear stress.
The repulsion and
stacking-fault tension act locally through the full, nonlinear $W_0/W$ repulsion. The applied stress $\sigma$ drives both partials equally [see Supplemental Material (SM)~\cite{sm}, Sec.~S1], whereas
the quenched disorder stresses $F_1$ and $F_2$ also drive their relative motion.  We
compare the free pair with a frozen-width control, $y_{1,2}=u\pm W_0/2$
(with mean position $u=(y_1+y_2)/2$), on the same disorder (numerical details in SM~\cite{sm}, Sec.~S2).

The disorder stresses are projections of one random-alloy stress field onto
the two partial Burgers vectors $\hat{\bm b}_{p,1}$ and $\hat{\bm b}_{p,2}$, with the anisotropic correlations of the
size-misfit model~\cite{GeslinRodney2021,GeslinRidaRodney2021,Rida2022}.
Each has variance
$\Delta_0$, and their normalized cross-correlation at separation $W$ in the
glide direction,
$
C(W)=\langle F_1(x,y)\,F_2(x,y-W)\rangle / \Delta_0
$,
follows from the stress correlator of that model (SM~\cite{sm}, Sec.~S3).
At coincident positions $C(0)=\hat{\bm b}_{p,1}\cdot\hat{\bm b}_{p,2}
=\cos60^\circ=1/2$, so the width is never unforced, $C$ decays over a length
$\xi_C$ ($1.35a$ for edge character) and is close to zero at the separation
$W_0=10a$ used below.
Lengths are measured in units of the regularization width $a$ of that
stress field, $a\simeq0.23\,a_{\rm lat}\approx0.8$~\AA\ in fcc~\cite{GeslinRodney2021}.

For the self-interaction we first use the long-range kernel
$\widehat{\mathcal{J}[y]}(q)=-c|q|\hat y(q)$ where hat denotes the Fourier transform, $c$ is a constant prefactor, and $q$ the wave number along the line direction.  It places the frozen
problem in the depinning class of contact lines and crack fronts, whose
exponents are known~\cite{Rosso2002,Duemmer2007}, so the control has an
answer fixed in advance. It also makes the ratio of partial coupling to
line elasticity a well-defined number, which with a local line tension
would carry $\ln(L/r_0)$ in the prefactor. The physical self-interaction of a dislocation
is instead proportional to $q^2\ln(1/q)$, and the partials also interact
nonlocally; we return to that kernel below.

\begin{figure}[t]
\centering
\includegraphics[width=\columnwidth]{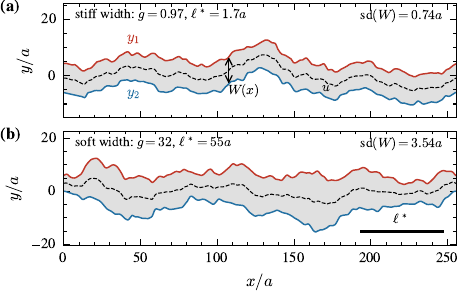}
\caption{The free stacking-fault width at the depinning threshold
($|q|$ kernel, $W_0=10a$, $\Delta_0=0.3$, $L=256a$, one realization).
Partials $y_1$ (upper) and $y_2$ (lower); shaded, the stacking fault of
local width $W(x)$; dashed, the center of mass $u$.  \textbf{(a)}~Stiff width: the partials move almost as one line.
\textbf{(b)}~Soft width: the bar marks $\ell^\ast=55a$, below which the
partials are pinned largely independently.}
\label{fig:model}
\end{figure}

Linearizing, the width obeys
\begin{equation}
\partial_t\delta W=\mathcal{J}[\delta W]-\kappa\,\delta W
+(F_1-F_2)+\ldots,
\label{eq:dW}
\end{equation}
with $\delta W=W-W_0$ and $\kappa=2\gamma/W_0$.  The restoring term makes the width a massive
mode.  Balancing $c|q|$ against $\kappa$ defines its elastic crossover
length $\ell^\ast=c/\kappa$: below $\ell^\ast$ width fluctuations are
governed by elasticity and the partials respond largely independently,
above it the width is slaved to the local disorder
(Fig.~\ref{fig:model}).  
The multistability
of the slaved width is measured by
$g=\sqrt{2\Delta_0[1-C(W_0)]}/(\kappa\xi_C)$, the typical gradient of the
disorder stress on the width over its stiffness, with $\xi_C$ the
glide-direction decorrelation length. For $g\gtrsim1$ the local width
equilibrium has multiple branches.

The coupling in Eqs.~\eqref{eq:1}--\eqref{eq:2} is monotone: the stress on each partial is
non-decreasing in the position of the other, and the kernel couples
distinct points positively.  Middleton's no-passing property therefore
holds~\cite{Middleton1992} and the threshold is unique (SM~\cite{sm}, Sec.~S4).  Together with
the periodicity of the disorder in the glide direction, with period
$L_y$, it yields a terminating test: relaxed at fixed $\sigma$, the pair
either arrests or advances every site by more than $L_y$, after which it
can never arrest.  
Bisection on this test gives the exact threshold $\sigma_c$ and critical configuration of each realization~\cite{RossoKrauth2001} (SM~\cite{sm}, Sec.~S5).
With periodic disorder the threshold statistics depend on $L_y$ relative to
the transverse extent of the line, which grows as $L^{\zeta}$ with the
roughness exponent $\zeta$~\cite{BolechRosso2004}. We therefore scale
$L_y\propto L^{0.385}$, the long-range value \cite{Rosso2002}, so that
finite-size scaling is at fixed aspect ratio.\\

\emph{Critical point}---We measure two exponents, both on the exact
critical configurations and thresholds: the roughness exponent $\zeta$
from the size dependence of the root-mean-square width of the center of
mass,
\begin{equation}
w^2(L)=\overline{\big\langle[u(x)-\langle u\rangle_x]^2\big\rangle_x},
\label{eq:wL}
\end{equation}
and the correlation-length exponent $\nu$ from the sample-to-sample
standard deviation of the threshold, $\mathrm{sd}(\sigma_c)\sim L^{-1/\nu}$.  
The free and frozen problems are solved on the same disorder realizations,
so we fit the differences $\Delta\zeta=\zeta_{\rm free}-\zeta_{\rm frozen}$
and $\Delta\nu$ directly, and sample-to-sample fluctuations common to both cancel.  All intervals quoted are 95\% bootstrap intervals: the fit is
repeated on realizations drawn with replacement at each size, each free and
frozen pair being drawn together.
We use five couplings,
$g=0.97$--$32$ ($\ell^\ast=1.7a$--$55a$) at $W_0=10a$ and $\Delta_0=0.3$,
with $L=64a$--$2048a$ and 100 realizations per size.

\begin{figure}[t]
\centering
\includegraphics[width=\columnwidth]{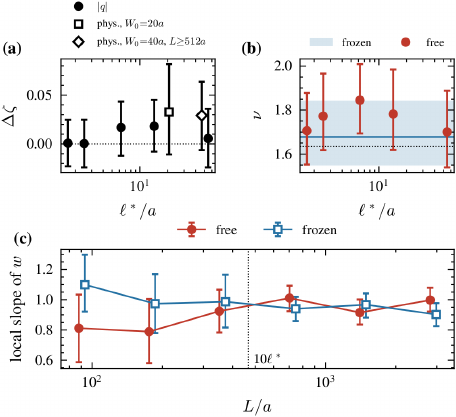}
\caption{The critical point is unchanged. 
\textbf{(a)}~$\Delta\zeta$ between free and frozen width against $\ell^\ast$, with 95\% intervals. 
Filled: $|q|$ kernel, five couplings, $L=64a$--$2048a$.  
Open: physical kernel, $W_0=20a$ (fit over $L=64a$--$2048a$) and $W_0=40a$ (fit over $L=512a$--$4096a$).  
\textbf{(b)}~$\nu$ of the free $|q|$ problem against $\ell^\ast$;
shaded, the frozen value, common to all couplings; dotted, $1/(1-0.388)$.
\textbf{(c)}~Local slopes of $w(L)$ (with 95\% intervals) for the physical kernel at $W_0=40a$: below
$L\approx10\ell^\ast$ (dotted) the free problem's slopes lie below the frozen
ones; above, the two coincide.}
\label{fig:critical}
\end{figure}

At every coupling the two problems share their critical point
(Fig.~\ref{fig:critical}): $\Delta\zeta$ lies between $0.000$ and $0.018$,
each 95\% interval containing zero with a half-width of about $0.03$,
and the free and frozen $\nu$ agree, with differences compatible with
zero.  This holds across a thirtyfold range of $g$, through the regime
where the slaved width is strongly multistable and its fluctuations
reach a third of $W_0$ [Fig.~\ref{fig:model}(b)]. The frozen exponents are $\nu=1.68 \pm 0.15$,
consistent with the statistical-tilt value ($1/(1-\zeta)=1.63$ for the long-range $\zeta$), and an
effective $\zeta=0.44 \pm 0.02$ that lies above the long-range value
$0.388$~\cite{Rosso2002}.  This offset is a property of the alloy
disorder at accessible sizes, not of the model: with uncorrelated
random-bond disorder the same simulations return $\zeta=0.381 \pm 0.021$
and $\nu=1.63$, while other disorder types give effective $\zeta$ between
$0.34$ and $0.48$ (SM~\cite{sm}, Sec.~S6).  The
paired differences are immune to this spread because both free and frozen-width problems share it.

The result follows from the massive character of the width.  On scales
beyond $\ell^\ast$ the width is slaved, and integrating it out leaves a
single line with renormalized disorder and mobility and a memory kernel
decaying on the time scale $1/\kappa$, but no new long-wavelength field.
Multistability does not overturn this: for $g\gtrsim1$ the slaved width
is multivalued and the effective force on $u$ is cusped and history
dependent, but at depinning the renormalized disorder correlator develops
a cusp anyway beyond the Larkin length~\cite{Chauve2001,LeDoussal2002,Rodney2024, wiese2022theory}.
Microscopic cusps change the bare disorder, not the fixed point, provided
the dynamics stay monotone, which they do.\\

\begin{figure}[t]
\centering
\includegraphics[width=\columnwidth]{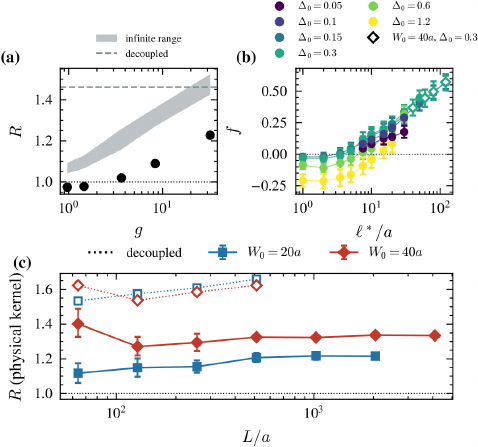}
\caption{The threshold shifts. 
\textbf{(a)}~$R$ against $g$ at $L=2048a$ ($|q|$ kernel); shaded, infinite-range model over its spring constants; dashed, decoupled limit $R_{\rm dec}$.
\textbf{(b)}~Normalized enhancement $f=(R-1)/(R_{\rm dec}-1)$ against $\ell^\ast/a$ at $L=512a$, for $W_0=10a$ (filled circles) and $W_0=40a$ at $\Delta_0=0.3$ (open diamonds).
Points are omitted where the width reaches the bounds $0.1W_0\le W\le W_0+40a$, within which the $W_0/W$ repulsion is evaluated, in more than 10\% of bisection trials or critical configurations (SM~\cite{sm}, Secs.~S2 and~S7).
\textbf{(c)}~Physical two-partial kernel at $\Delta_0=0.005$: $R$ against $L$ for $W_0=20a$ and $40a$ (Ni- and Cu-like dissociation), converged from $L=512a$ on.}
\label{fig:threshold}
\end{figure}

\emph{Threshold}---The stacking-fault width is therefore irrelevant in
the renormalization-group sense.  That does not make it irrelevant to the
threshold: the critical stress is not universal, and an irrelevant
mode may still move it.  We compare thresholds through
$
R=\sigma_c^{\rm free}/\sigma_c^{\rm frozen}
$.
At the largest size the free width lowers the threshold by
about 2\% when it is stiff and raises it by up to 23\% when it is soft
[Fig.~\ref{fig:threshold}(a)], with uncertainties of about $\pm1\%$. The
ratio no longer changes with $L$ beyond $L\approx512a$.
The upper reference is the decoupled
limit $\gamma\to0$, two partials in correlated disorder with no
interaction, for which $R_{\rm dec}=1.46$.

The coupling $g$ does not control the shift.  At fixed $g$, $R$ depends on
the disorder strength: at $g=8.34$ it is $1.107$ for $\Delta_0=0.15$ and
$1.012$ for $\Delta_0=0.6$, with disjoint intervals.  What grows
systematically is the normalized enhancement
$f=(R-1)/(R_{\rm dec}-1)$ as the width softens, i.e.\ with $\ell^\ast$,
from slightly negative for $\ell^\ast\lesssim3$--$10a$, depending on the
disorder, to about $0.5$ at $\ell^\ast\approx80a$ [Fig.~\ref{fig:threshold}(b)].
Measured in units of $a$, $\ell^\ast$ brings $\Delta_0=0.1$--$0.3$ and
$W_0=10a$ and $40a$ onto one curve within about $0.05$, but the weakest
disorder lies below it, as does the strongest at small $\ell^\ast$
(SM~\cite{sm}, Sec.~S7).

Both branches have a simple origin.  Below $\ell^\ast$ the partials
respond to their own stresses.  The frozen pair feels only their average,
of variance $\Delta_0[1+C(W_0)]/2$, while each free partial feels
$\Delta_0$, and stronger disorder pins harder. 
A soft width gives the pair access to that stronger pinning.
A stiff width offers instead a local relaxation channel: one partial
advancing while the other holds. It lowers the threshold by 1--5\%, more
the stronger the disorder and almost independently of $\gamma$
(SM~\cite{sm}, Sec.~S7).

The stability of the pinned pair shows where the threshold is
decided.  At $\sigma_c$ the lowest eigenvalue of its stability matrix
$-\partial\dot y_i/\partial y_j$ vanishes, along a mode localized over a
length $\xi$ set by the disorder ($37a$ at $\Delta_0=0.05$ to $5a$ at
$1.2$), with a makeup independent of $L$ (SM~\cite{sm},
Sec.~S8).  Its width content
$2P_W=\|\delta y_1-\delta y_2\|^2/(\|\delta y_1\|^2+\|\delta y_2\|^2)$, from
$0$ for a rigid translation to $1$ for one partial alone, is a single
function of $\ell^\ast/\xi$.  The enhancement $f$ is not: at
$\ell^\ast/\xi\approx3.4$ it ranges from $0.36$ to $0.03$ across the
disorder strengths, and it keeps growing with $\ell^\ast$ far beyond
$\xi$.  The threshold shift is therefore not set by the composition of the
instability that ends pinning.

An infinite-range version of the same model, in which each site couples to
the mean position and the width is eliminated site by site, gives a ratio
$R_{\rm MF}$ that grows steadily with $g$, to about $1.5$ at $g=32$
[shaded in Fig.~\ref{fig:threshold}(a); SM~\cite{sm}, Sec.~S9].
There the sites are independent,
$\sigma_c$ is an average of local stability limits, and every additional
width branch adds a limit~\cite{Fisher1985,NarayanFisher1992}: the width
can only create barriers.  Mean field therefore misses the softening
branch entirely, overestimates $R-1$ by a factor of $2$ at $g=32$ and
nearly ten at $g=3.65$, and has the wrong sign at $g\lesssim1.5$: it
attributes to multistability what is a crossover-length effect.
Models in which the partials interact only through their mean separation
fail in the opposite way: the partials decouple progressively and $R$
rises with $L$ toward $R_{\rm dec}$ (SM~\cite{sm}, Sec.~S2).\\

\emph{Physical kernel}---For a dislocation in isotropic elasticity, the
self-interaction of a partial 
\begin{equation}
K_s(q)=\frac{2C_s}{a_c}\Big[\frac1{a_c}-|q|K_1(|q|a_c)\Big],
\label{eq:kernel1}
\end{equation}
and the interaction between partials 
\begin{equation}
K_{12}(q)=2C_{12}\frac{|q|K_1(|q|W_0)}{W_0},
\label{eq:kernel2}
\end{equation}
follow from the line-integral energy (SM~\cite{sm}, Sec.~S10),
where $K_1$ denotes a modified Bessel function and $a_c$ a core cutoff.  
The prefactors $C_s=\mu b_pK_{ii}/4\pi$ and $C_{12}=\mu b_pK_{12}/4\pi$ are
set by the shear modulus $\mu$, the partial Burgers vector $b_p$, and the
isotropic energy factors $K_{ii}$ of one partial and $K_{12}$ of the pair,
which depend on the dislocation character [SM~\cite{sm}, Sec.~S1.3].
The self
kernel behaves as $C_sq^2\ln(2/|q|a_c)$ at small $q$, 
and the mutual kernel equals the pointwise, $q$-independent coupling of
Eqs.~\eqref{eq:1}--\eqref{eq:2} at $q=0$ but decays for $|q|W_0\gtrsim1$.
Elasticity now fixes the parameters that were free
before: $\gamma W_0=2C_{12}=1.27\,C_s$ for an edge dislocation, and the
width's crossover length, from the balance of its reduced elasticity
$K_s-K_{12}(0)+K_{12}(q)$ against $\kappa$, is $\ell^\ast/W_0=1.04$ at
$W_0=20a$ and $1.17$ at $W_0=40a$, close to unity for any realistic
dissociation.  These widths correspond to edge dislocations in Ni
($W_0\approx1.7$~nm) and Cu ($3.3$~nm)~\cite{HirthLothe1982}.  Both
kernels couple distinct points positively in real space, but with the
nonlinear $W_0/W$ repulsion the on-site coupling of the partials turns
negative once the fault opens by about $0.5\%$, so no-passing is not
guaranteed. 
However, thresholds from different initial states agree, and ordered
configurations stay ordered to within $3\times10^{-7}a$ (SM~\cite{sm}, Sec.~S4).

The $q^2\ln q$ kernel places the line in the short-range depinning class up
to logarithmic corrections, and the frozen-width case indeed shows an effective
$\zeta\simeq0.94$--$1.00$, between the long- and
short-range values.  
The critical point is again unchanged, with $\Delta\zeta=0.03\pm0.05$ at
$W_0=20a$.  At $W_0=40a$ the local slopes of the free problem lie below the
frozen ones up to $L\approx10\ell^\ast$ and coincide with them beyond, where
$\Delta\zeta=0.03\pm0.04$ and $\Delta\nu=0.14\pm0.19$
[Fig.~\ref{fig:critical}(c)].

The threshold shift is larger than with the $|q|$ kernel.  At
$\Delta_0=0.005$, $R=1.215 \pm 0.016$ for $W_0=20a$ and
$1.334 \pm 0.004$ for $W_0=40a$ at the largest sizes, unchanged from $512a$ 
on [Fig.~\ref{fig:threshold}(c)], and changing by at most $0.035$ when the
core cutoff is varied from $a$ to $2.5a$.  Here $f=0.33$ and $0.54$, above the
$|q|$-kernel values at the same $\ell^\ast$ (about $0.2$ and $0.4$): the decay of $K_{12}$ lets the
partials decouple below $W_0$, which the pointwise coupling of Eqs.~\eqref{eq:1}--\eqref{eq:2}
never allows.

To place real alloys, we calibrate $\Delta_0$ twice.  The mean-square
resolved shear stress of the size-misfit model~\cite{GeslinRodney2021} gives a site-level $\tau_{\rm rms}\approx0.85$--$1.15$~GPa
for CrMnFeCoNi, CrFeCoNi and CrCoNi, against measured 0\,K CRSS of
$168$--$225$~MPa~\cite{Kawamura2021,Li2022NiCoCr}.  The frozen
physical-kernel line reproduces this ratio,
$\sigma_c/\sqrt{\Delta_0}\approx0.20$--$0.24$, for
$\Delta_0\approx0.001$--$0.0025$ in units of $(C_s/a)^2$, while converting
$\tau_{\rm rms}$ with the elastic value of $C_s$ gives
$\Delta_0\approx0.003$--$0.005$ (SM~\cite{sm},
Sec.~S11).  Over $\Delta_0=0.0025$--$0.01$, $R=1.14$--$1.22$
at $W_0=20a$, decreasing toward weaker disorder, and $1.30$--$1.34$ at
$W_0=40a$, nearly independent of it.  Width fluctuations thus raise the
athermal CRSS by roughly $10$--$20\%$ for narrow and $30$--$35\%$ for
wide dissociation, relative to a rigid dissociated dislocation in the
same disorder.\\

\emph{Discussion}---An internal coordinate that is massive and locally
coupled leaves the depinning fixed point unchanged, however multistable it
is microscopically, but it is not irrelevant to the threshold.  It
shifts the threshold through two competing effects, local relaxation
when it is stiff and decoupling of the constituents below its elastic
crossover length when it is soft, and in a dissociated dislocation the
second wins.  Neither is a multistability effect, and the infinite-range
and mean-separation reductions misrepresent both.  An internal degree of
freedom that supports long-lived, extended structures, such as Bloch lines
in magnetic domain walls~\cite{Herranen2017,Skaugen2022,HonkanenLaurson2026},
is not bound by the fixed-point argument, which may explain why internal
structure has such different consequences in the two systems.

For alloy strength the conclusion is that rigid-dissociation theories~\cite{Leyson2010,Varvenne2016} omit a contribution of order $10$--$35\%$ to
the athermal CRSS, growing with the dissociation width.  The equilibrium
width also enters through the disorder the pair averages over, and the
two effects should not be conflated.  
One caution applies beyond this system: effective exponents at accessible
sizes depend on the correlations of the disorder at the level of $\pm0.05$
even when the universality class does not, which bears on the exponent
spread reported in atomistic simulations~\cite{Esfandiarpour2022}.

The model keeps the partials single valued and the disorder that of
isotropic size misfit, and works at zero temperature.  Modulus misfit,
short-range order, cross-slip, which the alloy disorder assists
(SM~\cite{sm}, Sec.~S12), and thermal activation, where
sub-threshold multistability may change barriers even where it does not
move the athermal threshold, are natural steps for future work.\\

\begin{acknowledgments}
\emph{Acknowledgments}---%
M.J.A. acknowledges support from the Research Council of Finland
(Grant No.~361245) and the Finnish Cultural Foundation. 
T.J.M. acknowledges support from the Research Council of Finland
(Grant No.~130265).
The authors acknowledge the computational resources provided by the Aalto University School of Science ``Science-IT'' project.
\end{acknowledgments}

%

\end{document}


\begin{center}
{\large\bfseries Supplemental Material for
``An internal mode shifts the depinning threshold of dissociated
dislocations without changing universality''}\\[6pt]
Mikko J. Alava and Tero J. M\"akinen
\end{center}
\vspace{1em}

\setcounter{equation}{0}
\setcounter{figure}{0}
\setcounter{table}{0}
\setcounter{section}{0}
\setcounter{secnumdepth}{3}
\renewcommand{\theequation}{S\arabic{equation}}
\renewcommand{\thefigure}{S\arabic{figure}}
\renewcommand{\thetable}{S\arabic{table}}
\renewcommand{\thesection}{S\arabic{section}}
\renewcommand{\thesubsection}{\thesection.\arabic{subsection}}
\makeatletter\renewcommand{\p@subsection}{}\makeatother

Equation, figure and table numbers prefixed by S refer to this
Supplemental Material, unprefixed numbers refer to the main text. 
Section~\ref{sm:cryst} gives the crystallography and elastic prefactors,
Sec.~\ref{sm:numerics} the numerical setup, Sec.~\ref{sm:disorder} the
disorder field, and Secs.~\ref{sm:nopassing} and~\ref{sm:threshold}
no-passing and the threshold protocol.  Sections~\ref{sm:exponents}--\ref{sm:calib}
give the data behind the results of the main text: exponents, threshold
scans, the stability analysis, the infinite-range model, the physical
kernel and the calibration of the disorder strength.
Section~\ref{sm:crossslip} records a by-product of the disorder statistics:
alloy disorder assists cross-slip.

\section{Crystallography and elastic prefactors}
\label{sm:cryst}
\subsection{Glide force on a partial}
\label{sm:pk}

The Peach--Koehler force per unit length on a dislocation with Burgers
vector $\bm b_p$ and line direction $\hat{\bm\xi}$ is
$(\bm\sigma\cdot\bm b_p)\times\hat{\bm\xi}$.  Only its component along
the glide direction $\hat{\bm s}=\hat{\bm n}\times\hat{\bm\xi}$ enters.
With $\bm t=\bm\sigma\cdot\bm b_p$,
\begin{equation}
(\bm t\times\hat{\bm\xi})\cdot(\hat{\bm n}\times\hat{\bm\xi})
=(\bm t\cdot\hat{\bm n})|\hat{\bm\xi}|^2
-(\bm t\cdot\hat{\bm\xi})(\hat{\bm\xi}\cdot\hat{\bm n}),
\end{equation}
and since $\hat{\bm\xi}\perp\hat{\bm n}$,
\begin{equation}
f_i=\hat{\bm n}\cdot\bm\sigma\cdot\bm b_{p,i}.
\label{eqS:pk}
\end{equation}
The line direction drops out.  Writing $\bm b_{p,i}=b_p\bm s_i$, the
force separating the partials is
\begin{equation}
f_1-f_2=b_p\,\hat{\bm n}\cdot\bm\sigma\cdot(\bm s_1-\bm s_2)
=-b_p\,\tau_E,
\label{eqS:escaig}
\end{equation}
using $\bm s_1-\bm s_2=-\hat{\bm e}$ (Table~\ref{tabS:cryst}), so the
width is driven by the Escaig stress alone.  For a Schmid stress
referred to the full Burgers vector,
$\bm\sigma=\tau(\hat{\bm b}\otimes\hat{\bm n}+\hat{\bm n}\otimes\hat{\bm b})$,
Eq.~(\ref{eqS:pk}) gives
$f_i=b_p\tau\,\hat{\bm b}\cdot\bm s_i=b_p\tau\cos30^\circ=b\tau/2$
for either character, since $b_p/b=1/\sqrt3$.  This justifies the
common $\sigma$ in Eqs.~(1)--(2).  It does not hold for a stress
referred to the glide direction of a mixed dislocation, for which the
width acquires a deterministic offset.

\subsection{Conventions}

We take $\bm b=\frac{a_{\rm lat}}{2}[110]$ (with $a_{\rm lat}$ the lattice constant), common to the two $\{111\}$ planes
containing it, and the Shockley pairs of Table~\ref{tabS:cryst}.  Each
pair sums to $\bm b$, lies in its plane, and has
$\hat{\bm b}_{p,1}\cdot\hat{\bm b}_{p,2}=\tfrac12$.  The magnitudes are
$b=a_{\rm lat}/\sqrt2$ and $b_p=a_{\rm lat}/\sqrt6$.  The Escaig direction is
$\hat{\bm e}_i=\hat{\bm n}_i\times\hat{\bm b}$ and for a screw it
coincides with the glide direction, which is why the separation of a
screw lies across $\bm b$.

\begin{table}[h]
\caption{Slip-system conventions.  The second row is the $(001)$
mirror image of the first, with $\hat{\bm n}_2$ taken in the sign for
which $\hat{\bm e}_2$ is the mirror image of $\hat{\bm e}_1$ ($\tau_E$
is invariant under the joint flip).}
\label{tabS:cryst}
\begin{ruledtabular}
\begin{tabular}{lcccc}
 & $\hat{\bm n}_i$ & $\bm b_{p,1}$ & $\bm b_{p,2}$ & $\hat{\bm e}_i$\\
\hline
primary & $\tfrac1{\sqrt3}[\bar111]$ & $\tfrac a6[211]$ &
$\tfrac a6[12\bar1]$ & $\tfrac1{\sqrt6}[\bar11\bar2]$\\
cross-slip & $\tfrac1{\sqrt3}[1\bar11]$ & $\tfrac a6[21\bar1]$ &
$\tfrac a6[121]$ & $\tfrac1{\sqrt6}[\bar112]$\\
\end{tabular}
\end{ruledtabular}
\end{table}

\subsection{Elastic prefactors and partial spacing}
\label{sm:prefactors}

For two parallel straight partials with Burgers vectors $\bm b_{p,1}$,
$\bm b_{p,2}$ and line direction $\hat{\bm\xi}$, the interaction force per
unit length is $\mu b_p^2K_{12}/(2\pi W)$ with
\begin{equation}
K_{12}=(\hat{\bm b}_{p,1}\cdot\hat{\bm\xi})(\hat{\bm b}_{p,2}\cdot\hat{\bm\xi})
+\frac{(\hat{\bm b}_{p,1}\times\hat{\bm\xi})\cdot(\hat{\bm b}_{p,2}\times\hat{\bm\xi})}{1-\nu_{\rm P}}
=\frac{2-\nu_{\rm P}-2\nu_{\rm P}\cos2\theta}{4(1-\nu_{\rm P})},
\label{eqS:K12}
\end{equation}
$\theta$ being the angle between the full Burgers vector and the line.
Balancing it against the stacking-fault energy $\Gamma$ gives the
equilibrium spacing~\cite{HirthLothe1982}
\begin{equation}
W_0=\frac{\mu b_p^2}{8\pi\Gamma}\,\frac{2-\nu_{\rm P}}{1-\nu_{\rm P}}
\left[1-\frac{2\nu_{\rm P}\cos2\theta}{2-\nu_{\rm P}}\right].
\label{eqS:W0}
\end{equation}
Dividing by $b_p$, as for every term of Eqs.~(1)--(2), turns this force into the
repulsion $\gamma W_0/W$ with $\gamma W_0=2C_{12}=\mu b_pK_{12}/2\pi$
(stress$\times$length). The self-interaction prefactor $C_s=\mu b_pK_{ii}/4\pi$
contains the energy factor of a single partial,
$K_{ii}=\cos^2\beta+\sin^2\beta/(1-\nu_{\rm P})$, with $\beta$ the angle between its
Burgers vector and the line. The partials of an edge dislocation lie at $\pm30^\circ$
from edge, so $K_{ii}=\tfrac14+\tfrac34/(1-\nu_{\rm P})=1.37$ and $K_{12}=0.87$ at
$\nu_{\rm P}=0.33$, which gives $\gamma W_0=1.27\,C_s$. For screw partials
$K_{ii}=1.12$ and $K_{12}=0.38$.

Note that the two partials have equal line tension only for pure edge and
pure screw character, so a common elastic prefactor in Eqs.~(1)--(2) would
not be justified for mixed character.

\subsection{Where real alloys sit}
\label{sm:gphys}

Eq.~(\ref{eqS:W0}) with $\nu_{\rm P}=0.33$ gives edge spacings of about $1.7$~nm for Ni
($\mu=76$~GPa, $\Gamma=125$~mJ\,m$^{-2}$~\cite{CarterHolmes1977}), $3.3$~nm for Cu
($48$~GPa, $45$~mJ\,m$^{-2}$~\cite{CarterRay1977}) and about $8$~nm for CrMnFeCoNi
($\mu\approx80$~GPa~\cite{Laplanche2015}, $\Gamma=30\pm5$~mJ\,m$^{-2}$ measured on single
crystals~\cite{Kawamura2021}), i.e.\ $W_0\approx20a$, $40a$ and $100a$.
The width crossover length follows
from the physical kernel of Sec.~\ref{sm:kernel}, not from a free
$\gamma$: with the reduced width elasticity and $a_c=1.5a$,
$\ell^\ast/W_0=0.88$, $1.04$, $1.17$ and $1.32$ at $W_0=10a$, $20a$,
$40a$ and $100a$.  Because $W_0$ and $\ell^\ast$ are both set by
$\mu b_p^2/\Gamma$, the ratio depends on the material only through a
logarithm, and every dissociated fcc dislocation sits at
$\ell^\ast/W_0\approx1$--$1.3$.  The disorder strength is calibrated in
Sec.~\ref{sm:calib}.

\section{Numerical setup}
\label{sm:numerics}

\subsection{Integration and units}

\emph{Unit of length.}  Lengths are measured in units of the Gaussian
regularization width $a$ of the microelasticity model of the alloy
stress field~\cite{GeslinRodney2021,GeslinRidaRodney2021}.  The
correlators $\Psi_L$ and $\Psi_T$ of Sec.~\ref{sm:disorder} are the
closed forms for that regularization with $u=r/a$, so this unit is fixed
by the disorder model.  Matching the continuum mean-square stress to the
lattice sum gives $a=0.23\,a_{\rm lat}$ for fcc~\cite{GeslinRodney2021}, about $0.83$~\AA\ for $a_{\rm lat}=3.6$~\AA.  Lengths quoted in
units of $a$ are therefore roughly four times smaller than the same
numbers in lattice constants.\\

\emph{Integration.}  Equations~(1)--(2) are integrated on $N_x=L/\mathrm{d}x$
points with $\mathrm{d}x=0.5a$, as in Ref.~\cite{Rida2022}, and periodic
boundary conditions in $x$.  The elastic kernel is evaluated spectrally.
Time stepping is explicit Euler, with $\Delta t=0.02$ for the $|q|$
kernel ($c=1$ sets the stress unit) and $\Delta t=0.2$ for the physical
kernel of Sec.~\ref{sm:kernel}. In every run the Euler map is checked to
be monotone, $\Delta t\,|\partial v_i/\partial y_i|<1$, which also makes
the reached metastable states independent of $\Delta t$.\\

\emph{Disorder field.}  The field is tabulated on a grid of spacing $\mathrm{d}y=0.25a$,
interpolated linearly to $y_i(x)$, and periodic in the glide direction with period
$L_y=48a\,(L/64a)^{0.385}$, rounded up to a multiple of $8a$. For the $|q|$ kernel this
fixes $L_y/L^\zeta$ across sizes, as finite-size scaling with periodic disorder requires.
Doubling the prefactor changes $\zeta$ by less than $0.01$ and $\nu$ within its interval
(Table~\ref{sm:exponents}). The same $L_y(L)$ is used for the physical kernel of Sec.~\ref{sm:kernel}, whose effective
roughness exponent is larger, so there $L_y$ falls from about $60$ to about $5$ times the
width $w(L)$ of the critical configuration between $L=64a$ and $4096a$. The free and
frozen problems share $L_y$, and $R$ does not change between $L=512a$ and $4096a$
(Table~\ref{tabS:kernel}).
Each
realization is normalized so that the stress on one partial has variance
exactly $\Delta_0$.  This removes the sample-to-sample scatter of the
field variance, which at $L=512a$ would contribute about $0.5\%$ to the
relative scatter of $\sigma_c$, against $7\%$ measured, so its effect on
$\nu$ is negligible.
For the stability analysis of Sec.~\ref{sm:stability} the same
tabulated field is interpolated instead with a periodic cubic B-spline,
which makes the landscape smooth, so that stability is lost at generic
folds rather than at the kinks of the linear interpolation.  On 100
matched realizations per size the cubic-spline thresholds are
$0.4$--$0.5\%$ higher in both problems, and $R$ is unchanged to within
$0.001$.\\

\emph{Repulsion.}  In the $W_0/W$ term the width is clipped to
$0.1W_0\le W\le W_0+40a$.  
The clip keeps the dynamics monotone, so thresholds remain exact, but where a bound is reached the repulsion differs from Eqs.~(1)--(2). How often this happens is reported in Sec.~\ref{sm:scan}, as is dissociation, in which the pair has no threshold (Sec.~\ref{sm:exact}).\\

\emph{Initial states and the frozen problem.}  Runs start from flat partials
at $\pm W_0/2$ with a jitter of $0.1a$.  The frozen-width control
integrates the center of mass alone,
\begin{equation}
\partial_t u=\mathcal{J}[u]+\sigma
+\tfrac12\big[F_1(x,u+W_0/2)+F_2(x,u-W_0/2)\big],
\label{eqS:frozen}
\end{equation}
with the width exactly $W_0$, on the same disorder realization as the
free pair.  Every free/frozen comparison is made seed by seed.  Since the
stacking-fault stress vanishes identically at $W=W_0$, the frozen problem is
independent of $\gamma$ and is computed once per $(L,\text{realization})$.\\

\emph{Parameters.}  Unless stated otherwise $W_0=10a$, $\Delta_0=0.3$ and
the edge character is chosen.  The five couplings used in the main simulations are
$\gamma=3.0$, $2.0$, $0.8$, $0.35$ and $0.0913$, giving $g=0.97$, $1.46$,
$3.65$, $8.34$ and $32$ and $\ell^\ast=c/\kappa=W_0/2\gamma=1.7a$,
$2.5a$, $6.3a$, $14.3a$ and $54.8a$.  At $W_0=10a$ the partials are
nearly decorrelated, $C(W_0)\simeq-0.04$.

\subsection{Modeling choices} \label{sm:choices}

\emph{Nonlinear repulsion.}  The $W_0/W$ form of the repulsion is kept
rather than linearized.  Under symmetric noise it rectifies the width
outward, which a harmonic well does not.\\

\emph{Pointwise versus averaged coupling.}  In some line-tension and discrete
implementations the partials interact only through their mean
separation~\cite{Vaid2022}. To test this we replaced $W(x)$ in the repulsion of
Eqs.~(1)--(2) by its line average $\bar W=\langle y_1-y_2\rangle_x$, so that every site
feels the same stacking-fault stress $\pm\gamma(W_0/\bar W-1)$. With this coupling
nothing restores the width locally: the partials
decouple progressively with $L$, 
cross each other in a fraction of critical configurations that grows with $L$ (from none at $L=64a$ to 16 of 20 at $L=512a$), and $R$ rises toward the decoupled limit
(Sec.~\ref{sm:scan}).\\

\emph{Elastic kernel.}  The $|q|$ kernel is a tractable idealization
chosen because its depinning exponents are known and because it makes the
ratio of partial coupling to line elasticity a well-defined number.  The
physical kernel, with $q^2\ln(1/q)$ self-elasticity and a nonlocal mutual
interaction, is treated in Sec.~\ref{sm:kernel}.

\section{Disorder statistics}
\label{sm:disorder}

\subsection{Correlation between two slip systems}

For randomly placed dilatation centers in an isotropic medium the
stress field is traceless, and the only isotropic form of its
correlation tensor is
$\langle\sigma_{ij}\sigma_{kl}\rangle=\mathcal{A}(\delta_{ik}\delta_{jl}
+\delta_{il}\delta_{jk}-\tfrac23\delta_{ij}\delta_{kl})$.  The resolved
shear stress on a slip system $(\bm s,\bm n)$ is
$\tau=s_i\sigma_{ij}n_j$.  For two systems with $\bm s\perp\bm n$ and
$\bm s'\perp\bm n'$ the trace term drops out, and the normalized
correlation is
\begin{equation}
C=(\bm s\cdot\bm s')(\bm n\cdot\bm n')+(\bm s\cdot\bm n')(\bm n\cdot\bm s').
\label{eqS:Cgen}
\end{equation}
Monte Carlo over $6\times10^4$ centers reproduces Eq.~(\ref{eqS:Cgen})
for randomly chosen pairs of slip systems.  The model is an isotropic
random distribution of size misfits.  Modulus disorder, elastic
anisotropy and chemical short-range order are not covered by it.

\subsection{Partial--partial correlator}
\label{sm:CW}

\begin{figure}[h]
\centering
\includegraphics[width=\textwidth]{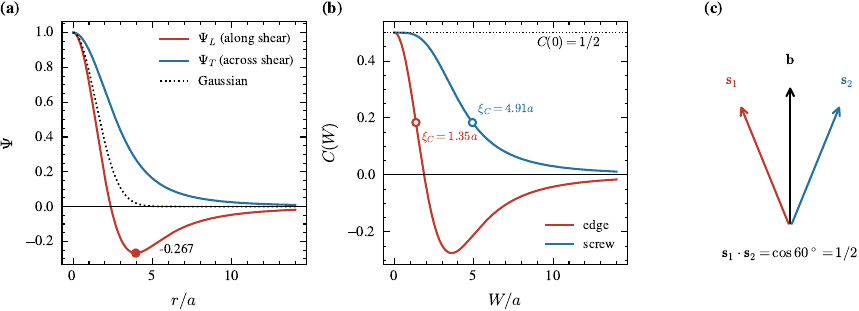}
\caption{Disorder statistics of a dissociated dislocation. 
\textbf{(a)}~The longitudinal and transverse correlators $\Psi_L$ and $\Psi_T$, with a Gaussian form for comparison.
Note the algebraic $u^{-3}$ tails and the negative lobe of $\Psi_L$, which reaches $-0.267$. 
\textbf{(b)}~The partial--partial correlator $C(W)$, Eq.~(\ref{eqS:CW}), for edge and screw characters, with the ceiling $C(0)=1/2$ and the $1/e$ lengths $\xi_C$ (these differ by a factor $3.6$ between the characters) marked. 
\textbf{(c)}~The two Shockley slip directions at $60^\circ$, whose scalar product is the ceiling in panel (b).}
\label{figS:corr}
\end{figure}

We write the in-plane shear-stress correlation tensor at separation $r$
along a unit vector $\bm e$ as
\begin{equation}
K_{ij}(r)=\Psi_L(r)\,e_ie_j+\Psi_T(r)\,(\delta_{ij}-e_ie_j),
\qquad \Psi_L(0)=\Psi_T(0)=1,
\label{eqS:K}
\end{equation}
with $\Psi_L$ and $\Psi_T$ the longitudinal and transverse functions of
the microelasticity model~\cite{GeslinRidaRodney2021,Rida2022}.  Each
partial feels $F_i=\bm s_i\cdot\bm\tau$, so for partials separated by
$W$ along $\bm e$
\begin{equation}
C(W)=(\bm s_1\cdot\bm e)(\bm s_2\cdot\bm e)\,\Psi_L(W)
+\big[\bm s_1\cdot\bm s_2-(\bm s_1\cdot\bm e)(\bm s_2\cdot\bm e)\big]\Psi_T(W).
\label{eqS:CW}
\end{equation}
At zero separation $C(0)=\bm s_1\cdot\bm s_2=\cos60^\circ=1/2$,
independent of everything else (direct Monte Carlo gives $0.503$ for
the Shockley pair on $(11\bar1)$, Fig.~\ref{figS:corr}).  For the correlators used here,
$C(W)\le1/2$ at all $W$.

Two consequences follow.  The width forcing $F_1-F_2$ has variance
$2\Delta_0[1-C(W)]\ge\Delta_0$ at every separation, including zero, so
the internal mode is never unforced.  The center-of-mass forcing
$(F_1+F_2)/2$ has variance $\Delta_0[1+C(W)]/2\le0.75\,\Delta_0$.  What
switches the width off at small $W_0$ is the divergence of the
stiffness $\kappa=2\gamma/W_0$, not a vanishing of the forcing.

\subsection{Edge and screw characters}
\label{sm:edgescrew}

\begin{figure}[h]
\centering
\includegraphics[width=\textwidth]{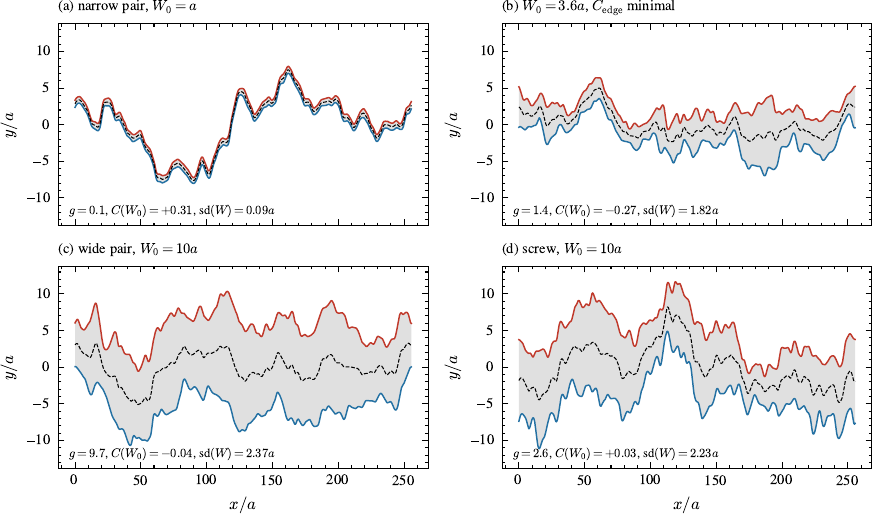}
\caption{Critical configurations in the four regimes the correlator distinguishes, one realization each, $L=256a$, $\Delta_0=0.3$.  $\gamma W_0=3$ is fixed, so panels (a)--(c) follow the path of
an alloy series. 
\textbf{(a)}~$W_0=a$: the stiff pair moves as one line.
\textbf{(b)}~$W_0=3.6a$, where $C_{\rm edge}$ reaches its minimum $-0.27$ and the width forcing is maximal. 
\textbf{(c)}~$W_0=10a$, $C\simeq0$: the partials are pinned almost independently.  
\textbf{(d)}~A screw at the same $W_0$ and $\gamma W_0$: because edge and screw exchange which correlator governs which channel, Eq.~(\ref{eqS:Cchar}), the screw width is smoother.}
\label{figS:configs}
\end{figure}

For an edge the separation lies along $\bm b$, so
$\bm s_1\cdot\bm e=\bm s_2\cdot\bm e=\cos30^\circ$.  For a screw it
lies across $\bm b$, so $\bm s_{1,2}\cdot\bm e=\pm\sin30^\circ$.
Equation~(\ref{eqS:CW}) then gives
\begin{equation}
C_{\rm edge}=\tfrac34\Psi_L-\tfrac14\Psi_T,\qquad
C_{\rm screw}=-\tfrac14\Psi_L+\tfrac34\Psi_T.
\label{eqS:Cchar}
\end{equation}
The edge inherits the negative lobe of $\Psi_L$.  $C_{\rm edge}$ crosses
zero near $W=1.9a$ and reaches $-0.27$ at $W\simeq3.6a$, where the
width forcing is $27\%$ above that of independent partials.  The $1/e$
decorrelation lengths are $\xi_C=1.35a$ for the edge and $4.91a$ for
the screw.  Because $\int_0^\infty\Psi_L\,\mathrm{d}u=0$ while
$\int_0^\infty\Psi_T\,\mathrm{d}u=15\sqrt\pi/8$, one channel always has
vanishing spectral weight at zero wavenumber.  For edges this is the
center-of-mass channel, and for screws the width channel, so
long-wavelength width fluctuations are suppressed for screws but not
for edges (Fig.~\ref{figS:configs}).

\subsection{Spectral synthesis}
\label{sm:synthesis}

The two-component field $\bm\tau(x,y)$ is generated by two-dimensional fast Fourier transform.  In Fourier space an isotropic vector field decomposes into
channels longitudinal and transverse to $\bm k$,
\begin{equation}
\hat K_{ij}(\bm k)=S_L(k)\,\hat k_i\hat k_j
+S_T(k)\,(\delta_{ij}-\hat k_i\hat k_j),
\label{eqS:spec}
\end{equation}
and we set
$\hat{\bm\tau}(\bm k)=\sqrt{S_L}\,\eta_L\hat{\bm k}
+\sqrt{S_T}\,\eta_T\hat{\bm k}_\perp$, with $\eta_{L,T}$ independent
complex Gaussian variables of the appropriate Hermitian symmetry.  The
field is realizable because $S_L,S_T\ge0$.  
Averaging over the angle $\phi$ between $\bm k$ and $\bm r$, with
$\langle\cos^2\!\phi\,e^{ikr\cos\phi}\rangle_\phi=[J_0(kr)-J_2(kr)]/2$ and
$\langle\sin^2\!\phi\,e^{ikr\cos\phi}\rangle_\phi=[J_0(kr)+J_2(kr)]/2$, where $J_n$ is the
Bessel function of the first kind of order $n$, gives the real-space functions of
Eq.~(\ref{eqS:K}),
\begin{align}
\Psi_L(r)&\propto\int_0^\infty\!\frac{k\,\mathrm{d}k}{2\pi}
\Big[S_L(k)\,\frac{J_0(kr)-J_2(kr)}{2}+S_T(k)\,\frac{J_0(kr)+J_2(kr)}{2}\Big],\\
\Psi_T(r)&\propto\int_0^\infty\!\frac{k\,\mathrm{d}k}{2\pi}
\Big[S_L(k)\,\frac{J_0(kr)+J_2(kr)}{2}+S_T(k)\,\frac{J_0(kr)-J_2(kr)}{2}\Big].
\end{align}
In practice the real-space
functions are the input: with $u=r/a$,
\begin{align}
\Psi_L(u)&=-\frac{30}{u^3}\Big[\sqrt\pi\Big(1-\frac{12}{u^2}\Big)
\mathrm{erf}\frac u2+\frac{12+u^2}{u}\,e^{-u^2/4}\Big],\\
\Psi_T(u)&=\frac{15}{u^3}\Big[\sqrt\pi\Big(1-\frac{6}{u^2}\Big)
\mathrm{erf}\frac u2+\frac6u\,e^{-u^2/4}\Big],
\end{align}
the closed forms for a Gaussian-regularized dilatation center~\cite{GeslinRidaRodney2021}, set to $1$ for $u<0.06$.  The tensor
$K_{ij}$ is tabulated on the $(x,y)$ grid, transformed, and projected
onto $\hat{\bm k}$ to obtain $S_L$ and $S_T$. The small negative
values that discretization produces are clipped to zero.  Each
partial feels $F_i=\bm s_i\cdot\bm\tau$, so the auto- and
cross-correlations follow from the projection rather than being
imposed.  The measured glide-direction cross-correlation agrees with
Eq.~(\ref{eqS:CW}) to better than $0.005$.

\subsection{Error of the single-field treatment}
\label{sm:singlefield}

\begin{figure}[h]
\centering
\includegraphics[width=0.5\textwidth]{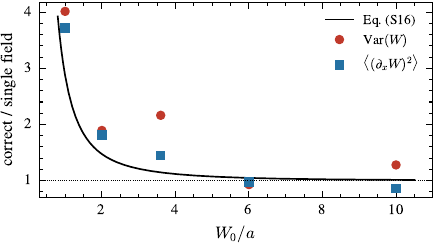}
\caption{Ratio of internal-mode fluctuations between the correct
two-field disorder and the single-field treatment, for
$\mathrm{Var}(W)$ and $\langle(\partial_xW)^2\rangle$, against
Eq.~(\ref{eqS:varratio}) with no adjustable parameter; $L=256a$, four
realizations at the critical configuration.  The error is largest at narrow separation and has
gone by $W_0\simeq10a$.}
\label{figS:single}
\end{figure}

Setting $F_2\equiv F_1$, a single scalar field sampled at the two
partial positions, recovers the conventional treatment.  Since the
width forcing has variance $2\Delta_0[1-C(W_0)]$, the linear response
of Eq.~(3) gives
\begin{equation}
\frac{\mathrm{Var}(W)|_{\rm correct}}{\mathrm{Var}(W)|_{\rm single}}
=\frac{1-C_{\rm correct}(W_0)}{1-C_{\rm single}(W_0)}.
\label{eqS:varratio}
\end{equation}
At $W_0=a$ this predicts 2.87, against 4.01 and 3.72 measured for $\mathrm{Var}(W)$ and $\langle(\partial_xW)^2\rangle$ (Fig.~\ref{figS:single}). Equation~\eqref{eqS:varratio} captures the trend but underestimates the effect up to $W_0 \approx 3.6a$, where $C_{\rm edge}$ is most negative and the measured ratios are still 0.93 and 0.98. From $W_0 \approx 6a$ on, both are close to one within the scatter of four realizations (1.28 and 0.86 at $W_0 = 10a$). The single-field treatment therefore underestimates width fluctuations by up to a factor of four at narrow separation, and the error is negligible at the $W_0 \geq 10a$ used in the main text.

\section{No-passing}
\label{sm:nopassing}

\emph{Monotonicity.}  Middleton's argument~\cite{Middleton1992}
requires that the force on each degree of freedom be non-decreasing in
the positions of all others.  The kernel $\mathcal{J}$ couples distinct
points of the same line positively.  The stacking-fault term couples
the two lines, and
\begin{equation}
\frac{\partial}{\partial y_2}\Big[\gamma\Big(\frac{W_0}{W}-1\Big)\Big]
=\frac{\partial}{\partial y_1}\Big[-\gamma\Big(\frac{W_0}{W}-1\Big)\Big]
=\frac{\gamma W_0}{W^2}>0 .
\end{equation}
Each disorder stress depends only on its own partial's position.  The
condition therefore holds for as long as $W>0$.\\

\emph{Tests.}  We prepared ordered pairs of configurations and advanced
them under identical disorder and drive, recording the minimum
separation over space and time.  Three sweeps were run:
\begin{itemize}
\item edge, six values of $g$ from $0.97$ to $8.34$, at
$\sigma/\sigma_c\in\{0.8,1.0,1.5\}$;
\item edge, five values of $g$ from $9.73$ to $32.5$, at
$\sigma/\sigma_c\in\{0.9,0.95,1.0,1.05,1.2\}$;
\item screw, four values of $g$ from $0.97$ to $8.34$, at the first set
of drives.
\end{itemize}
The minimum separation was non-negative in all $55$ configurations.
The margin exceeds $0.15a$ for $g\le3.65$, falls to
$+1.5\times10^{-4}a$ near $g=5.8$, and recovers to $\sim10^{-2}a$ by
$g=24$, so the dip is not a trend toward violation.  Screw
configurations keep margins of $0.30a$ or more throughout.
The ordered pairs are the flat initial state and the same state with
both partials advanced by $2a$, evolved for a time $200$.\\

\emph{Physical kernel.}  In Eqs.~(5)--(6), the self kernel has
real-space couplings $C_s/(s^2+a_c^2)^{3/2}$ and the mutual kernel
$C_{12}/(s^2+W_0^2)^{3/2}$, both positive.  In the model, however,
the mutual interaction is the nonlinear pointwise term of
Eqs.~(1)--(2) plus the nonlocal correction
$2C_{12}[h(q)-1/W_0^2]$ of Sec.~\ref{sm:kernel}, linearized at $W_0$, whose
real-space value at zero distance is negative.  The coupling between the
two partials at the same $x$, $\gamma W_0/W^2$ plus that value, is positive
at $W=W_0$ but changes sign once the fault opens beyond
$W\simeq1.0067W_0$ ($W_0=20a$) or $1.0044W_0$ ($W_0=40a$), which happens at a
large fraction of sites.  
Middleton's condition is therefore not guaranteed for this kernel, and neither are
the uniqueness of the threshold, the termination argument of Sec.~\ref{sm:exact}, or the practice,
also justified there by no-passing, of starting each bisection trial from the last
pinned configuration.  
We tested them directly at $L=64a$ and $256a$, $\Delta_0=0.005$, in 8
realizations for each of $W_0=20a$ and $40a$ at each size.  
The threshold was located four ways. Three use the bisection
of the main simulations, begun from different states: flat partials, a flat state with
the fault opened to $1.5W_0$, and partials displaced by independent random amounts of
standard deviation $W_0/4$. These three values coincide. In the fourth, every trial
starts afresh from the flat initial state, and the result differs from the others by
at most $4\times10^{-4}$ in relative terms, inside the bisection tolerance of
$10^{-3}$, in every realization at both sizes.
Ordered pairs of configurations, prepared from the same three kinds of
start state and advanced by $2a$, were evolved for a time $2000$ at
$\sigma/\sigma_c=0.9$, $1.0$ and $1.05$ (288 runs), during which the fault opened
to up to $2.16\,W_0$, deep in the range where the on-site coupling is negative.
The configurations crossed by at most $2.4\times10^{-7}a$ (at $L=256a$,
$W_0=40a$; at $L=64a$ the smallest separation was $-3\times10^{-14}a$, zero to
round-off).  Any violation of no-passing is therefore negligible, and the
thresholds are unaffected.

\section{Locating the threshold}
\label{sm:threshold}

\label{sm:exact}

Monotonicity and the periodicity of the disorder in the glide direction
give a test for pinning with no test time.  Relaxed at fixed $\sigma$
from a configuration $Y_0$, the pair either arrests or advances every site
of both partials by more than $L_y$.  In the second case the current state
lies everywhere ahead of $Y_0+L_y$, itself the start of a trajectory of
the same system, so by no-passing it stays ahead and advances without
bound.  Arrest is declared when the largest velocity falls below
$10^{-6}$ and stays below it for three times the slowest elastic
relaxation time.  Bisection on this test, with relative bracket width
$10^{-3}$ and every trial started from the last pinned configuration
(which no-passing permits), gives the exact $\sigma_c$ and the critical
configuration of each realization, as for the single long-range string~\cite{RossoKrauth2001,Rosso2002}.  
We performed several checks on the main simulation code: the
field is periodic to $10^{-14}$; the Euler map is monotone; a critical
configuration evolved for a time $20L$ at the lower bracket end moves by
less than $10^{-4}a$; at the upper end it depins; the frozen threshold is
independent of $\gamma$ to the bracket width; results are reproducible.
The test decides in every realization except dissociating ones.  The
stacking-fault attraction on each partial is bounded by $\gamma$
($0.8\gamma$ under the clip of Sec.~\ref{sm:numerics} at $W_0=10a$), so when
the trailing partial alone is harder to depin than the leading one by more
than $2\gamma$, no stress depins the pair as a whole: the leading partial
runs away while the trailing one stays pinned, the relaxation never
terminates, and the realization has no threshold of the pair.  This is the
model's counterpart of the unbounded expansion of a stacking-fault ribbon.
In the main simulations it occurs in one realization, at $g=32$ and $L=64a$,
where the two partial thresholds differ by $2.34\gamma$, and at no other
coupling or size. We have excluded that realization (Table~\ref{tabS:production}).  In the scans it occurs only at the softest points (Sec.~\ref{sm:scan}).  For the physical kernel the termination
argument rests on no-passing, which there is supported numerically rather
than guaranteed (Sec.~\ref{sm:nopassing}).

A criterion based on a fixed observation time is not used: arrest
times diverge at threshold, so such a criterion places $\sigma_c$ too low by
an amount that depends on the observation time, while the terminating test
has no such parameter.

\section{Critical exponents}
\label{sm:exponents}

\begin{figure}[h]
\centering
\begin{tikzpicture}
    \node[inner sep=0pt] (img) at (0, 0) {\includegraphics[width=\textwidth]{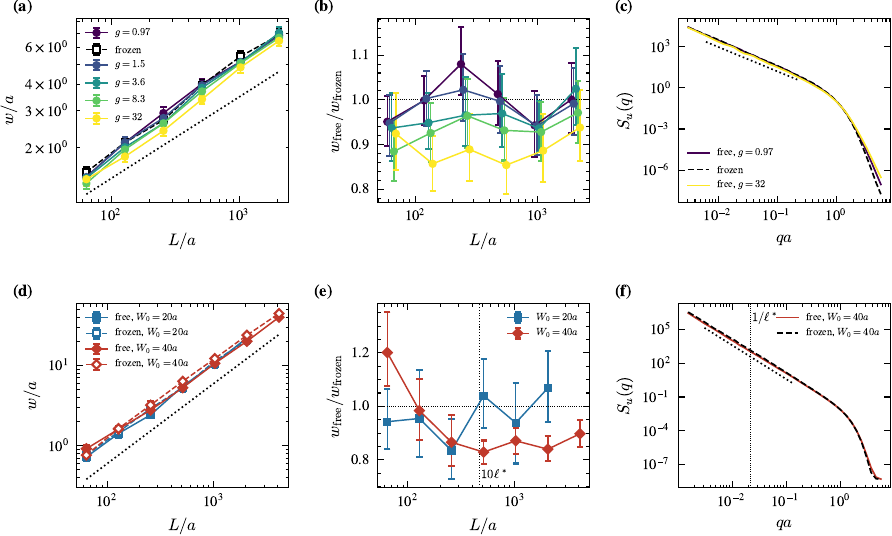}};
    \node[inner sep=0pt, rotate=32] (text1) at (-0.25\textwidth, 0.16\textwidth) {\scriptsize $L^{0.388}$};
    \node[inner sep=0pt, rotate=-21] (text2) at (0.34\textwidth, 0.22\textwidth) {\scriptsize $q^{-1.78}$};
    \node[inner sep=0pt, rotate=40] (text3) at (-0.25\textwidth, -0.14\textwidth) {\scriptsize $L^{1}$};
    \node[inner sep=0pt, rotate=-30] (text4) at (0.315\textwidth, -0.097\textwidth) {\scriptsize $q^{-3}$};
\end{tikzpicture}
\caption{Roughness of the critical configurations.
Top panels:~$|q|$ kernel ($W_0=10a$, $\Delta_0=0.3$, 100 realizations).
Bottom panels: physical kernel ($\Delta_0=0.005$).
Error bars are 95\% bootstrap intervals.
\textbf{(a)},\textbf{(d)}~Center-of-mass width $w(L)$, free (filled) and frozen (open, dashed).
Dotted line is $w \propto L^{0.388}$ in panel (a) and $w \propto L^{1}$ in panel (d) for reference.
\textbf{(b)},\textbf{(e)}~The paired ratio $w_{\rm free}/w_{\rm frozen}$, whose flatness at large $L$ is $\Delta\zeta=0$. The dotted vertical line in panel (e) is $10\ell^\ast$ for $W_0=40a$.
\textbf{(c)},\textbf{(f)}~Center-of-mass structure factor $S_u(q)$ at the largest size.
The dotted lines are $q^{-(1+2\zeta)}$ for $\zeta=0.388$ and $\zeta=1$. In panel (f) $q=1/\ell^\ast$ is marked.}
\label{figS:roughness}
\end{figure}

\emph{Estimators.}  $\zeta$ is obtained from $w(L)\sim L^{\zeta}$,
Eq.~(4), on the exact critical configurations, and
independently from the disorder-averaged structure factor
$S(q)\sim q^{-(1+2\zeta)}$ at the largest $L$, log-binned over
wavelength windows from $16a$ to $L/4$. On synthetic lines with
$\zeta=0.4$ the latter returns $0.397 \pm 0.021$.  $\nu$ is obtained
from $\mathrm{sd}(\sigma_c)\sim L^{-1/\nu}$.  Differences between the free
and frozen problems are fitted directly, $\Delta\zeta$ as the slope of
$\ln[w_{\rm free}/w_{\rm frozen}]$ against $\ln L$, with paired bootstrap
intervals as in the main text.  Figure~\ref{figS:roughness} shows the raw
data behind the fits: $w(L)$, the paired ratio $w_{\rm free}/w_{\rm frozen}$,
and $S_u(q)$ for both kernels.\\

\begin{table}[h]
\caption{Main simulation run, $|q|$ kernel, $W_0=10a$, $\Delta_0=0.3$,
$L=64a$--$2048a$, 100 realizations per size (99 at $g=32$, $L=64a$,
where one realization dissociates, Sec.~\ref{sm:exact}). Error estimates correspond to 95\% intervals.  The frozen
problem is common to all couplings: $\zeta=0.442 \pm 0.021$,
$\nu=1.68 \pm 0.15$.}
\label{tabS:production}
\begin{ruledtabular}
\begin{tabular}{cccccc}
$g$ & $\ell^\ast/a$ & $R(L=2048a)$ & $\Delta\zeta$ & $\nu_{\rm free}$ & $\Delta\nu$\\
\hline
$0.97$ & $1.7$  & $0.974\ [0.968,0.981]$ & $0.001\ [-0.024,0.025]$ & $1.71\ [1.55,1.90]$ & $0.03\ [-0.11,0.18]$\\
$1.46$ & $2.5$  & $0.978\ [0.970,0.986]$ & $0.000\ [-0.024,0.026]$ & $1.77\ [1.61,1.96]$ & $0.09\ [-0.07,0.26]$\\
$3.65$ & $6.3$  & $1.020\ [1.011,1.029]$ & $0.017\ [-0.012,0.045]$ & $1.85\ [1.70,2.03]$ & $0.17\ [-0.03,0.37]$\\
$8.34$ & $14.3$ & $1.089\ [1.079,1.100]$ & $0.018\ [-0.008,0.045]$ & $1.78\ [1.62,1.98]$ & $0.10\ [-0.10,0.33]$\\
$32$   & $54.8$ & $1.227\ [1.215,1.241]$ & $0.006\ [-0.026,0.035]$ & $1.70\ [1.56,1.88]$ & $0.02\ [-0.18,0.24]$\\
decoupled & $\infty$ & $1.461\ [1.447,1.476]$ & $-0.041\ [-0.071,-0.010]$ & $1.68\ [1.49,1.91]$ & $0.00\ [-0.25,0.26]$\\
\end{tabular}
\end{ruledtabular}
\end{table}

\emph{Disorder dependence of the effective exponents.}  The frozen problem
is a single long-range line, for which $\zeta=0.388$ and $\nu=1.63$~\cite{Rosso2002,Duemmer2007}.  Frozen-only runs at $L=64a$--$2048a$ with
100 realizations, Table~\ref{tabS:controls}, show that its effective
$\zeta$ at accessible sizes depends on the disorder: uncorrelated
random-bond disorder ($F=-\partial_yV$ of an uncorrelated potential)
recovers the known values, uncorrelated random-field disorder lies about
$0.02$--$0.03$ above them, the edge alloy disorder about $0.05$ above, and
the screw alloy disorder below by the $S(q)$ estimator.  
At depinning, random-bond and random-field disorder flow to the same
fixed point~\cite{wiese2022theory}, so all the disorders of
Table~\ref{tabS:controls} belong to one universality class.
The random-bond control, which recovers the known exponents, validates the simulations, and the spread among the others reflects disorder-specific corrections to scaling. Doubling the aspect-ratio prefactor of $L_y$ changes $\zeta$ by less than 0.01 and $\nu$ within its interval. The spread cancels in the paired differences.\\

\begin{table}[h]
\caption{Exponents of the frozen problem for different disorders, $|q|$ kernel.}
\label{tabS:controls}
\begin{ruledtabular}
\begin{tabular}{lccc}
disorder & $\zeta$ from $S(q)$ (four windows) & $\zeta$ from $w(L)$ & $\nu$\\
\hline
alloy, edge & $0.43$--$0.48$ & $0.442\ [0.422,0.464]$ & $1.68\ [1.54,1.83]$\\
alloy, edge, $L_y$ doubled & $0.45$--$0.47$ & $0.448\ [0.428,0.470]$ & $1.56\ [1.44,1.71]$\\
uncorrelated random field & $0.40$--$0.43$ & $0.408\ [0.390,0.426]$ & $1.81\ [1.63,2.02]$\\
uncorrelated random field, $L_y$ doubled & $0.41$--$0.43$ & $0.412\ [0.392,0.432]$ & $1.80\ [1.64,1.97]$\\
uncorrelated random bond & $0.386$--$0.394$ & $0.381\ [0.360,0.401]$ & $1.63\ [1.49,1.79]$\\
alloy, screw & $0.33$--$0.35$ & $0.403\ [0.384,0.422]$ & $1.72\ [1.57,1.89]$\\
\end{tabular}
\end{ruledtabular}
\end{table}

\emph{Crossover at small $L$.}  At sizes comparable to $\ell^\ast$ the
partials are pinned largely independently, which lowers the effective
exponent of the free problem, as in the decoupled limit, which shows a negative
$\Delta\zeta$ (Table~\ref{tabS:production}, last row).  In the
main simulations this is visible only at $L\le512a$
($\Delta\zeta\approx-0.05$ on average) and disappears by $L=2048a$. With
the physical kernel at $W_0=40a$ it is resolved explicitly
(Sec.~\ref{sm:kernel}).

\section{Threshold scans}
\label{sm:scan}

All scans use $L=512a$, at which $R$ has saturated in the main simulations, and 50 realizations per point. Readings (iv) and (v) below come from an earlier, smaller run with 20 realizations per point, at $L=512a$ for (iv) and $L=64a$--$512a$ for (v).
The normalized enhancement
is $f=(R-1)/(R_{\rm dec}-1)$, with $R_{\rm dec}$ the decoupled ratio at the
same $\Delta_0$ and $W_0$.
Dissociating realizations (Sec.~\ref{sm:exact}) are excluded.  They
occur only at the softest points, $\ell^\ast/W_0=20$ (2, 4 and 5 of 50 at
$\Delta_0=0.3$, $0.6$ and $1.2$) and $\ell^\ast/W_0=12$ (1 of 50 at
$\Delta_0=0.6$ and $1.2$), all of which are also affected by the clip.\\

\begin{table}[h]
\caption{Normalized enhancement $f$ against $\ell^\ast/W_0=c/2\gamma$,
$|q|$ kernel, $W_0=10a$ unless marked.  
Daggers mark entries at which the width reaches a bound of the clip $0.1W_0\le W\le W_0+40a$ of Sec.~\ref{sm:numerics} in more than $10\%$ of bisection trials or critical configurations. There the repulsion is not the $W_0/W$ of Eqs.~(1)--(2), and these entries are omitted from Fig.~3(b).
Bottom row:~decoupled ratio.}
\label{tabS:scan}
\begin{ruledtabular}
\begin{tabular}{ccccccccc}
$\ell^\ast/W_0$ & $\Delta_0=0.05$ & $0.1$ & $0.15$ & $0.3$ & $0.6$ & $1.2$ & $0.3$, $W_0=40a$\\
\hline
$0.1$  & --- & --- & $-0.02$ & $-0.03$ & $-0.09$ & $-0.21$ & ---\\
$0.35$ & --- & --- & $0.01$ & $-0.01$ & $-0.07$ & $-0.19$ & ---\\
$0.75$ & $0.045$ & $0.082$ & --- & --- & --- & --- & ---\\
$1$    & $0.080$ & $0.116$ & $0.15$ & $0.16$ & $0.05$ & $-0.04$ & $0.37$\\
$1.5$  & $0.127$ & $0.159$ & $0.19$ & $0.20$ & $0.14$ & $0.03$ & $0.45$\\
$2$    & $0.137$ & $0.210$ & $0.24$ & $0.24$ & $0.23$ & $0.08$ & $0.50$\\
$3$    & $0.177$ & $0.289$ & $0.30$ & $0.33$ & $0.32$ & $0.19^\dagger$ & $0.57$\\
$8$    & --- & --- & $0.49$ & $0.54^\dagger$ & $0.51^\dagger$ & $0.49^\dagger$ & ---\\
$20$   & --- & --- & $0.67^\dagger$ & $0.70^\dagger$ & $0.69^\dagger$ & $0.63^\dagger$ & ---\\
\hline
$R_{\rm dec}$ & $1.65$ & $1.62$ & $1.56$ & $1.46$ & $1.36$ & $1.25$ & $1.42$\\
\end{tabular}
\end{ruledtabular}
\end{table}

\emph{Readings.}  (i) At $\ell^\ast/W_0\lesssim0.35$, $R=0.988$, $0.985$,
$0.965$ and $0.947$ at $\Delta_0=0.15$, $0.3$, $0.6$ and $1.2$, nearly
independent of $\gamma$.  (ii) $f$ grows monotonically with $\ell^\ast$ in
every column.  (iii) At fixed $\ell^\ast/W_0$, $f$ falls at weak disorder
and is about twice as large at $W_0=40a$ as at $10a$.  At fixed
$\ell^\ast$ itself, in units of $a$, the data come together
[Fig.~3(b)]: $\Delta_0=0.1$--$0.3$ agree within about
$0.05$, and the $W_0=40a$ points fall on the $W_0=10a$ curve within about
$0.05$, so the $W_0$ dependence at fixed $\ell^\ast/W_0$ is that of the
fourfold larger $\ell^\ast$.  $\Delta_0=0.05$ lies lower by $0.04$--$0.15$,
increasingly so with $\ell^\ast$; $\Delta_0=0.6$ and $1.2$ lie lower at
small $\ell^\ast$, where the stiff-width softening pulls $R$ below one, and
approach the common curve for $\ell^\ast\gtrsim30$--$80a$.  The critical-mode
scan of Sec.~\ref{sm:stability} gives the same ordering
[Fig.~\ref{figS:collapse}(c)].  (iv) At fixed $g$, $R$ depends on
$\Delta_0$ ($g$ held fixed by $\gamma\propto\sqrt{\Delta_0}$):
$1.107$, $1.095$, $1.012$ at $g=8.34$ and $1.252$, $1.221$, $1.125$ at
$g=32$ for $\Delta_0=0.15$, $0.3$, $0.6$, so $g$ is not the control
parameter.  (v) With the mean-separation coupling of
Sec.~\ref{sm:choices} at $g=3.65$, $R=1.02$, $1.09$, $1.20$, $1.26$ at
$L=64a$, $128a$, $256a$, $512a$.\\

\emph{Clipping.}  In the main simulations no critical configuration reaches the lower bound
of the clip of Sec.~\ref{sm:numerics}, and bisection trials reach it only at $g=32$, in at most $5\%$ of
them ($L=2048a$). The upper bound, $W\ge W_0+40a$, is reached only at $g=32$, in the
critical configurations of 6, 5 and 3 realizations at $L=64a$, $128a$ and $256a$, and of
none from $512a$ on, where the widest fault stays below $W_0+34a$. Leaving those
realizations out lowers $R$ at $g=32$ by at most $0.013$, less than one standard error.
In the physical-kernel runs the lower bound is never reached, and the upper bound only at
$W_0=40a$, in the critical configurations of 1 of 50 realizations at $\Delta_0=0.01$
($L=512a$) and of 1 and 3 of 400 at $\Delta_0=0.005$ ($L=2048a$ and $4096a$), the widest
fault being $W_0+42a$. In the scans neither bound is reached at $\ell^\ast/W_0\le3$ for
$\Delta_0\le0.6$. The entries marked in Table~\ref{tabS:scan} and omitted from Fig.~3(b) are those at
which the lower bound is reached in $26$--$82\%$ of bisection trials, with partials
crossing in up to $22\%$ of critical configurations at $\ell^\ast/W_0=20$. Counting the
upper bound as well marks no further entries. In the independent critical-mode scan of
Sec.~\ref{sm:stability}, the points at which either bound is reached in more than $10\%$ of critical
configurations, or more than $2\%$ of realizations dissociate, form a subset of the
marked ones.

\section{Stability at the threshold}
\label{sm:stability}

\emph{Method.}  The dynamics of Eqs.~(1)--(2) are a
gradient flow, so the stability matrix $M_{ij}=-\partial\dot y_i/\partial
y_j$ of a pinned state, with $i,j$ running over the sites of both
partials, is symmetric, and the pinned state loses stability at $\sigma_c$
where its lowest eigenvalue vanishes.  The disorder field is interpolated
with a periodic cubic B-spline (Sec.~\ref{sm:numerics}), so the landscape
is smooth and the threshold is a saddle-node fold.  Thresholds are located
by following pinned branches to their folds, to a relative accuracy of
$10^{-10}$, for the free pair, the frozen problem and each partial alone
(the decoupled limit), and the lowest eigenvector $(\delta y_1,\delta
y_2)$ at the fold is obtained by dense diagonalization of $M$.  Its width
content is $2P_W=\|\delta y_1-\delta y_2\|^2/(\|\delta y_1\|^2+\|\delta
y_2\|^2)$.  Since the off-diagonal elements of $M$ are non-positive, the
lowest mode has no node, $\delta y_1\delta y_2\ge0$ everywhere, and
$0\le2P_W\le1$, from a rigid translation of the pair to one partial moving
alone.  Its extent is the participation length
$\xi=\mathrm{d}x\,(\sum_x\alpha)^2/\sum_x\alpha^2$ with
$\alpha=\delta y_1^2+\delta y_2^2$.\\

\emph{Runs.}  Two runs, both with the $|q|$ kernel.  The base run uses the
five couplings from the main simulations at $W_0=10a$ and $\Delta_0=0.3$, with
$L=64a$--$512a$ and 400 realizations per size.  The scan covers
$\ell^\ast=1a$--$200a$ at $W_0=10a$ and $\Delta_0=0.05$, $0.1$, $0.15$,
$0.3$, $0.6$ and $1.2$, and $\ell^\ast/W_0=1$--$3$ at $W_0=40a$ and
$\Delta_0=0.3$, at $L=256a$ and $512a$ with 100 realizations per point.
Both reproduce the thresholds of the main text: at $L=512a$ the base run
gives $R=0.983$, $0.984$, $1.016$, $1.075$ and $1.212$ for the five
couplings, against $0.984$, $0.982$, $1.023$, $1.084$ and $1.218$ in the
main simulations, and the scan reproduces Table~\ref{tabS:scan} within its
uncertainties.\\

\begin{figure}[h]
\centering
\includegraphics[width=\textwidth]{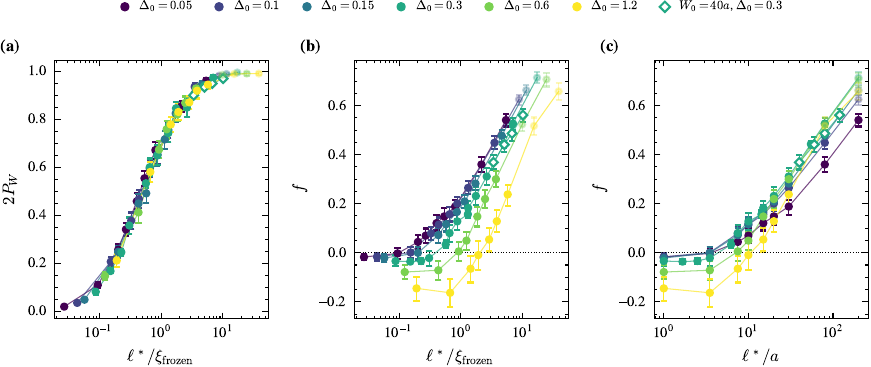}

\caption{Critical mode and threshold shift in the critical-mode scan ($|q|$ kernel, cubic-spline disorder, $L=512a$; $W_0=10a$ at six disorder strengths and $W_0=40a$ at $\Delta_0=0.3$; 100 realizations per point, 400 at the five couplings of the main simulations). Error bars correspond to 95\% bootstrap intervals over realizations.
\textbf{(a)}~Width content $2P_W$ of the critical mode against $\ell^\ast/\xi_{\rm frozen}$, with $\xi_{\rm frozen}$ the mean participation length of the critical mode of the frozen problem at the same disorder.
\textbf{(b)}~The normalized enhancement $f$ against the same variable.
\textbf{(c)}~$f$ against $\ell^\ast$ in units of $a$.
Faded symbols: points at which the width reaches a bound of the clipped repulsion (Sec.~\ref{sm:numerics}) in more than $10\%$ of critical configurations, or more than $2\%$ of realizations dissociate. The dissociated realizations are excluded from the means.}
\label{figS:collapse}
\end{figure}

\emph{Results.}  (i) The critical mode is localized.  In the base run $\xi$
is $8$--$11a$, almost independent of the coupling; it decreases slowly with
$L$ and saturates.  For the frozen problem it shrinks sevenfold with the
disorder, $\xi=37$, $23$, $17$, $11.5$, $8$ and $5a$ at
$\Delta_0=0.05$--$1.2$ ($L=512a$).  (ii) Its composition does not depend on
$L$ from $64a$ on: at $L=512a$, $2P_W=0.17$, $0.25$, $0.53$, $0.75$ and
$0.94$ for the five couplings, while $R$ still drifts by about $0.02$
between $128a$ and $512a$ at the intermediate couplings.  The threshold is
thus triggered by a local instability, and the drift of $R$ is the
finite-size approach of the thresholds.  (iii) Each realization has
$1.2$--$2.5$ instabilities, growing with $L$, in the last $3\%$ of stress
below $\sigma_c$, with the same $2P_W$ as the final one, so $2P_W$
characterizes the near-threshold instabilities rather than a single
event.  (iv) For the softest width the trailing partial carries the mode
in $66$--$75\%$ of realizations.  (v) Over the scan, $2P_W$ is a single
function of $\ell^\ast/\xi$, with $\xi$ that of the frozen problem at the
same disorder [Fig.~\ref{figS:collapse}(a)]: the families scatter about a
common curve with a root-mean-square (rms) deviation of $0.005$--$0.011$, against $0.12$ when
plotted against $\ell^\ast$.  (vi) The enhancement $f$ is not a function of
$\ell^\ast/\xi$ [Fig.~\ref{figS:collapse}(b), rms deviation $0.08$]: at
$\ell^\ast/\xi\approx3.4$, $f=0.36$, $0.31$, $0.21$ and $0.03$ at
$\Delta_0=0.05$, $0.3$, $0.6$ and $1.2$, and the sign change of $f$ falls
anywhere between $\ell^\ast/\xi=0.1$ and $2$.  Against $\ell^\ast$ in units
of $a$ the families come together [Fig.~\ref{figS:collapse}(c), rms
deviation $0.03$--$0.045$], as in Fig.~3(b), and $f$ keeps
growing far beyond $\xi$.  Since $a$, $\mathrm{d}x$ and $\mathrm{d}y$ were
not varied, these data do not separate the disorder correlation length
from the grid as the length in units of which $\ell^\ast$ orders $f$.

\section{Infinite-range model}
\label{sm:meanfield}

In the infinite-range model the elastic kernel is replaced by a
coupling of each site to the mean position of its line through a
spring of constant $k$,
\begin{equation}
\partial_t y_{a,i}=k(\bar y_a-y_{a,i})+\sigma
\pm\gamma\Big(\frac{W_0}{W_i}-1\Big)+F_a(i,y_{a,i}),
\qquad a=1,2,
\label{eqS:mf}
\end{equation}
with the frozen problem defined analogously from Eq.~(\ref{eqS:frozen}).
Since the spring acts on each partial separately, the width acquires
an additional stiffness $k$, small compared with $\kappa$ over the
range used.  Given the means, the sites are independent.  The
width at each site is eliminated by following its force balance
adiabatically as the drive is raised.  For $g\gtrsim1$ this proceeds by
branch jumps rather than smoothly.  The collective threshold is then
the average of the resulting local stability limits~\cite{Fisher1998,NarayanFisher1993}.  Thresholds of both problems are
located on the same realizations, with the same disorder statistics
and the same $\Delta_0$, $W_0$ and $\gamma$ as the resolved model,
but with $N_x=256$ sites and a disorder period $L_y=64a$, and by
bisection on the terminating test of Sec.~\ref{sm:exact} with a stress
tolerance of $10^{-6}$, as for the resolved model.

Over $g=0.97$--$32$ and $k=0.005$--$0.08$, with eight realizations at each
of the fifteen points (Table~\ref{tabS:meanfield}), $R_{\rm MF}$ runs from
$1.04$ to $1.52$.  It
grows monotonically with $g$ and more weakly with $k$.

\begin{table}[h]
\caption{Infinite-range threshold ratio $R_{\rm MF}$ against coupling $g$
and spring constant $k$, eight realizations per point, compared with the
spatially resolved $R$ at $L=2048a$ (Table~\ref{tabS:production}).}
\label{tabS:meanfield}
\begin{ruledtabular}
\begin{tabular}{ccccc}
$g$ & $k=0.005$ & $k=0.02$ & $k=0.08$ & resolved $R$\\
\hline
$0.97$ & $1.042$ & $1.053$ & $1.093$ & $0.974$\\
$1.46$ & $1.066$ & $1.082$ & $1.129$ & $0.978$\\
$3.65$ & $1.158$ & $1.197$ & $1.258$ & $1.020$\\
$8.34$ & $1.271$ & $1.311$ & $1.361$ & $1.089$\\
$32$   & $1.425$ & $1.475$ & $1.525$ & $1.227$\\
\end{tabular}
\end{ruledtabular}
\end{table}

\section{Physical two-partial kernel}
\label{sm:kernel}

\emph{Kernel.}  The self-energy of a regularized partial of core radius
$a_c$ and the interaction energy of two partials a distance $W_0$ apart,
expanded to second order in their displacements, give the kernels of
Eqs.~(5)--(6).  The self kernel follows from the same
cancellation of $K_0$ terms that produces the mutual kernel. It is
positive at all $q$, behaves as $C_sq^2[\ln(2/|q|a_c)-\gamma_E+\tfrac12]$
for $|q|a_c\ll1$ and saturates at $2C_s/a_c^2$.  (A form with $+q^2K_0$
and the opposite sign of the constant, which turns negative for
$|q|a_c\gtrsim1.1$, should not be used.)  The mutual kernel enters as a
nonlocal correction $2C_{12}[h(q)-1/W_0^2]$, $h=|q|K_1(|q|W_0)/W_0$, to the
pointwise coupling of Eqs.~(1)--(2), whose nonlinear $W_0/W$ form is kept.
With $C_s=1$ as the unit of stress$\times$length, $\gamma W_0=2C_{12}=
2C_sK_{12}/K_{ii}=1.27$ for edge partials, and $\Delta_0$ is in units
of $(C_s/a)^2$.  The decoupled limit sets both $\gamma$ and $C_{12}$ to
zero, and the frozen problem keeps the mutual kernel, which acts on the center of
mass.  The crossover length solves
$K_s(q)-2C_{12}[W_0^{-2}-h(q)]=2\gamma/W_0$ at $q=1/\ell^\ast$.\\

\begin{table}[h]
\caption{Physical kernel, $a_c=1.5a$: $R$ at $L=512a$, and in parentheses at the
largest size simulated ($2048a$ for $W_0=20a$, $4096a$ for $40a$), with 95\%
intervals; $R_{\rm dec}$ at $L=512a$ for the three $\Delta_0$.  Fifty realizations
per point, except $W_0=40a$, $\Delta_0=0.005$, which has 400 from $L=512a$ on.}
\label{tabS:kernel}
\begin{ruledtabular}
\begin{tabular}{cccccc}
$W_0$ & $\ell^\ast/W_0$ & $\Delta_0=0.0025$ & $0.005$ & $0.01$ & $R_{\rm dec}$\\
\hline
$20a$ & $1.04$ & $1.144$ & $1.207$ ($1.215\ [1.199,1.231]$) & $1.223$ & $1.70$, $1.66$, $1.57$\\
$40a$ & $1.17$ & $1.306$ & $1.325$ ($1.334\ [1.331,1.338]$) & $1.301$ & $1.66$, $1.62$, $1.52$\\
\end{tabular}
\end{ruledtabular}
\end{table}

\emph{Size and cutoff.}  At $\Delta_0=0.005$, $R$ is constant from $L=512a$ on
[Fig.~3(c), Table~S6]. Varying the core cutoff over $a_c=a$--$2.5a$ on the same 50
realizations changes $R$ by at most $0.01$ at $W_0=20a$ and by up to $0.035$ at
$W_0=40a$ (between $a_c=1.5a$ and $2.5a$ at $L=512a$).\\

\emph{Exponents.}  The frozen problem gives an effective $\zeta\simeq0.92$--$1.05$
at $L\le512a$, $0.955\ [0.920,0.990]$ over $L=64a$--$2048a$ at $W_0=20a$ and
$0.940\ [0.914,0.965]$ over $L=512a$--$4096a$ at $W_0=40a$, with $S(q)$ giving
$0.95$--$1.00$: between the long-range $0.388$ and the short-range $1.25$, as
expected for a $q^2\ln q$ kernel whose logarithmic corrections decay only as
$1/\ln L$.  At $W_0=20a$, $\Delta\zeta=0.012$, $0.023$ and $-0.014$ for the three
$\Delta_0$ at $L\le512a$, and $\Delta\zeta=0.033\ [-0.015,0.081]$ and
$\Delta\nu=-0.17\ [-0.47,0.10]$ over $L=64a$--$2048a$.  At $W_0=40a$,
$\Delta\zeta=-0.16$ to $-0.19$ over $L=64a$--$512a$, from a crossover: the local
slopes of the free problem are $0.81$, $0.79$, $0.92$, $1.01$, $0.92$ and $1.00$,
and of the frozen one $1.10$, $0.97$, $0.99$, $0.94$, $0.97$ and $0.90$, between
successive sizes from $64a$ to $4096a$.  With 400 realizations at each size from
$512a$ on, $\Delta\zeta=0.029\ [-0.006,0.066]$ and $\Delta\nu=0.14\ [-0.05,0.32]$
over $L=512a$--$4096a$, and $\Delta\zeta=0.022\ [-0.032,0.078]$ and
$\Delta\nu=0.05\ [-0.24,0.38]$ over $L=1024a$--$4096a$.  The ratio of free to
frozen threshold scatter is $0.85$, $0.90$, $0.99$ and $0.92$ at $L=512a$,
$1024a$, $2048a$ and $4096a$, constant within its $\pm5\%$ uncertainty.  Beyond
the crossover the center of mass of the free pair remains 10--15\% smoother
than the frozen line at every size, an amplitude rather than an exponent effect.

\section{Calibration of the disorder strength}
\label{sm:calib}

\emph{Stress amplitude.}  In the size-misfit model the mean-square
resolved shear stress is~\cite{GeslinRodney2021}
\begin{equation}
\langle\tau^2\rangle=\frac{v_{\rm at}\,\Delta\varepsilon^2\mu^2}
{30\pi^{3/2}a^3}\left(\frac{1+\nu_{\rm P}}{1-\nu_{\rm P}}\right)^2,
\label{eqS:tau2}
\end{equation}
with $\Delta\varepsilon^2$ the eigenstrain variance and $v_{\rm at}$ the
atomic volume. With the parameters of Ref.~\cite{Rida2022} it reproduces
their $\langle\tau^2\rangle=0.611$~GPa$^2$ for Al$_{50}$Mg$_{50}$.  With
$a=0.23a_{\rm lat}$ and misfit parameters $\delta=0.019$, $0.017$ and
$0.021$~\cite{Yin2020}, it gives $\tau_{\rm rms}\approx0.85$, $0.85$ and
$1.15$~GPa for CrMnFeCoNi, CrFeCoNi and CrCoNi, uncertain by about
$\pm40\%$ and excluding modulus misfit and short-range order.  Their
measured 0\,K CRSS, $168$, $200$ and $225$~MPa~\cite{Kawamura2021,Li2023CrFeCoNi,Li2022NiCoCr}, give $\tau_c/\tau_{\rm rms}\approx0.20$--$0.24$.

\begin{itemize}
    \item \emph{Route 1: strength ratio.}  Between $\Delta_0=0.0025$ and $0.01$ the frozen
physical-kernel threshold scales as $\sigma_c\propto\Delta_0^{0.66}$, the weak-pinning
law $\sigma_c\propto\Delta_0^{2/3}$ of a line with line
tension~\cite{Labusch1970,LarkinOvchinnikov1979}, so
$\sigma_c/\sqrt{\Delta_0}\approx0.27\,(\Delta_0/0.005)^{1/6}$ after
extrapolation to large $L$.  Matching $0.20$--$0.24$ gives
$\Delta_0\approx0.001$--$0.0025$.  The sixth-power dependence makes this
estimate sensitive and contributions to the real CRSS not contained in
$\tau_{\rm rms}$ push it lower.
    \item \emph{Route 2: elastic conversion.}  In physical units $C_s/a=\mu b_pK_{ii}/(4\pi a)\approx0.19\mu$,
about $15$--$17$~GPa for the three alloys with $\mu\approx80$--$90$~GPa~\cite{Laplanche2015,laplanche2018elastic},
so $\Delta_0=(\tau_{\rm rms}a/C_s)^2\approx0.003$--$0.005$.
\end{itemize}

The two routes bracket $\Delta_0\approx0.001$--$0.005$.  The $|q|$ kernel,
for which the conversion of $c$ is scale dependent, gives a consistent
picture: its frozen line has $\sigma_c\propto\Delta_0^{0.66}$ and
$\sigma_c/\sqrt{\Delta_0}\approx0.26\,(\Delta_0/0.3)^{1/6}$, matched at
$\Delta_0\approx0.06$--$0.2$ in units of $c^2$.

\section{Cross-slip stress statistics}
\label{sm:crossslip}

For a screw with $\bm b=\frac{a_{\rm lat}}{2} [110]$ able to glide on the two planes
of Table~\ref{tabS:cryst}, Eq.~(\ref{eqS:Cgen}) gives three results.
The Schmid stress $\hat{\bm n}_i\cdot\bm\sigma\cdot\hat{\bm b}$ and the
Escaig stress $\hat{\bm n}_j\cdot\bm\sigma\cdot\hat{\bm e}_j$ are
exactly uncorrelated, both within a plane and across the two.  This
follows because $\hat{\bm b}$ lies in both planes and is perpendicular
to both $\hat{\bm e}_j$.  The Escaig stresses on the two planes are
anti-correlated, with
$C=(\hat{\bm e}_1\cdot\hat{\bm e}_2)(\hat{\bm n}_1\cdot\hat{\bm n}_2)
+(\hat{\bm e}_1\cdot\hat{\bm n}_2)(\hat{\bm n}_1\cdot\hat{\bm e}_2)
=\tfrac19-\tfrac89=-\tfrac79$.

The sign of the widening force follows from Eq.~(\ref{eqS:escaig}).
From Table~\ref{tabS:cryst},
$\bm b_{p,1}-\bm b_{p,2}=-b_p\hat{\bm e}_i$ on both planes, with
partial~1 leading.  Which ordering is physical need not be determined,
because only the relative sign between the planes enters.  The $(001)$
mirror is a symmetry of the fcc crystal.  It fixes $\bm b$ and
$\hat{\bm\xi}=[110]$ and exchanges the two planes.  Being improper, it
maps the primary configuration onto the cross-plane configuration of
the opposite screw.  Recovering the correct screw negates both partial
Burgers vectors, and restoring an intrinsic fault then exchanges the
leading and trailing roles.  Each operation reverses
$\bm b_{p,1}-\bm b_{p,2}$, and together they leave it unchanged.  The
widening force therefore carries the same sign on both planes.

Friedel--Escaig cross-slip requires constriction on one plane and
re-dissociation on the other, and hence Escaig stresses of opposite
sign at the same point.  The anti-correlation makes that combination
the likely one.  Alloy disorder therefore assists cross-slip, through a
channel statistically independent of glide pinning.

%